# Ultra-low carrier density superconducting bolometers with single photon sensitivity based on magic-angle twisted bilayer graphene

G. Di Battista[1], K.C. Fong[2,3], A. Díez-Carlón[1], K. Watanabe[4], T. Taniguchi[5] and D. K. Efetov[1,6]*

1. Fakultät für Physik, Ludwig-Maximilians-Universität, Schellingstrasse 4, München 80799, Germany
2. Department of Physics, Harvard University, Cambridge, MA 02138, USA
3. Quantum Engineering and Computing Group, Raytheon BBN Technologies, Cambridge, MA 02138, USA
4. Research Center for Functional Materials, National Institute for Materials Science, 1-1 Namiki, Tsukuba 305-0044, Japan
5. International Center for Materials Nanoarchitectonics, National Institute for Materials Science, 1-1 Namiki, Tsukuba 305-0044, Japan
6. Munich Center for Quantum Science and Technology (MCQST), München, Germany

*E-mail : dmitri.efetov@lmu.de

**Abstract**
**The superconducting (SC) state of magic-angle twisted bilayer graphene (MATBG) shows exceptional properties, as it consists of an unprecedentedly small electron (hole) ensemble of only ~ $10^{11}$ carriers/cm$^{-2}$, which is five orders of magnitude lower than in traditional superconductors. This results in an ultra-low electronic heat capacity and kinetic inductance of this truly two-dimensional SC, and provides record-breaking key parameters for a variety of quantum sensing applications, in particular in thermal sensing and single photon detection (SPD), which traditionally exploit thermal effects in nanostructured superconducting thin films. In this work, we systematically study the interaction of the superconducting state of MATBG with individual light quanta. We discover full destruction of the SC state upon absorption of a single infrared photon even in a 16 μm$^2$ sized device, which showcases its exceptional bolometric sensitivity. Upon voltage biasing close to its critical current, we further show that this non-optimized device can be used as a SPD, whose click-rate is proportional to the number of absorbed photons, following Poissonian statistics. Our work offers insights into the MATBG-photon interaction and shows up pathways to use low-carrier density graphene-based superconductors as a new platform for developing revolutionary new quantum devices and sensors.**

Superconducting materials are at the heart of advanced technologies, as they are central active elements for modern quantum computing, quantum sensing and quantum metrology applications. In particular, nanopatterned superconducting thin films have gained significant attention for ultrasensitive photodetection[1–3], as these combine a low heat capacity and a sharp superconducting transition. When a photon is absorbed in such a device, it breaks Cooper pairs and generates quasiparticles above the superconducting gap, thereby introducing a change in impedance. Harnessing this mechanism, superconductor-based detectors, such as transition-edge sensors[4,5], superconducting nanowires[6–8], hot electron bolometers[9], kinetic-inductance detectors[10] and Josephson junctions[11–13] are among the best photodetectors for applications demanding high sensitivity, e.g. communication, radio astronomy[14], quantum network[15], and spectroscopy[16].

Novel two-dimensional superconductors offer a unique approach to single photon detection (SPD), due to their reduced electronic heat capacity and electron-phonon coupling, leading to a large temperature rise of the electron ensemble upon absorption of single photons[17–21]. More intriguingly, it has been recently discovered that the flat electronic band structure produced by stacking two layers of graphene twisted at a 'magic' angle leads to a novel superconducting phase[22,23]. The discovery has now expanded into an entire family of graphene-based superconductors[24–27], not only sparking intense investigations to understand the fundamental physics of their alleged unconventional superconducting states, but also prompting exploration of their potential applications[28–30]. As the intrinsic moiré superlattice of MATBG supports a record-low carrier density of $n \sim 10^{11}$ cm$^{-2}$, which is $\sim 5$ orders of magnitude lower than that of conventional superconductors (Fig. 1b), even a minute amount of quasiparticles generated by a single low-energy photon can significantly perturb the superconducting state. Specifically, in such a low-carrier density material, the substantial change in kinetic inductance induced by the absorbed photon could be readout using microwave resonators, opening a promising avenue to extend SPD across a broader spectral range.

In this study, we take the first step to develop a SPD based on superconducting MATBG and perform a proof-of-principle experiment to demonstrate for the first time the capability of detecting single-photons. We illuminate the device at mK temperatures with a highly attenuated 1550-nm laser source and monitor the induced photovoltage ($V_{ph}$), as shown in the schematic drawing of Fig. 1a.

**Results**

The optical image (inset of Fig. 1f) shows a typical device. The van der Waals stack consists of two graphene sheets rotated at a global twist-angle of $\sim 1.1°$ encapsulated into insulating hexagonal boron nitride (hBN) layers. The metallic graphite gate underneath the heterostructure is used to electrostatically tune the carrier concentration in the MATBG by applying an external gate voltage. Fig. 1d shows the four-terminal longitudinal resistance $R_{xx}$ of device A ($\theta = 1.04°\pm0.02$) as a function of the moiré filling factor $v$ (filling of electrons per moiré unit cell) for temperatures ranging from $T = 50$ mK up to $T = 6$ K. At electrostatic doping levels corresponding to the half-filling of the moiré unit cell ($v = -2$), we observe an insulating state flanked by a pronounced superconducting dome[22]. In Fig. 1e, the measurement of $R_{xx}$ vs. $T$ at the optimal doping of $v = -2.45$ reveals a superconducting transition with a normal state resistance of $\sim 10$ k$\Omega$ and critical temperature of $T_c \sim 2.8$ K, calculated as 50% of its normal state resistance (see Supplementary Information, section A for a complete transport characterization).

Fig. 1f plots the $I$–$V$ characteristic of the superconducting state of device A measured in a 4-terminal current-biased scheme at $T = 35$ mK. Here, we observe a clear hysteretic behavior with respect to the sweeping direction of the bias current $I_{bias}$, characterized by $\Delta I = I_c - I_r \sim 15$ nA, where $I_c$ and $I_r$ are the switching and retrapping current respectively (see Supplementary Information, section B). The hysteresis loops are ubiquitous in MATBG[31], potentially due to a current-induced self-heating hotspot[32,33] when the MATBG is in the normal state.

**Photoresponse measurements**

We can bias our device near the normal-superconductor transition to enable SPD. When the photon is absorbed, it breaks Cooper pairs and produces a voltage output. In order to prevent the detector to "latch" permanently into a stable resistive state where it no longer detects photons, we implement a self-reset circuitry (Fig. 1c). The circuit is constituted by a voltage divider with load resistor $R_2 \ll R_{res} + R_{MATBG}$. Here $R_{res}$ is a residual resistance (arising from the contact resistance and the metallic leads) and $R_{MATBG}$ is the 4-terminal resistance of the device active region, sketched as a variable resistor. The voltage bias scheme maintains a constant voltage across the source and drain contacts of the device ($V_{bias}$). In this way, the increase of resistance induced by the switching of the MATBG detector in the normal state, diverts part of the current into the load resistor $R_2$ reducing the current flowing in the detector and consequently the Joule heating, analogously to an electrothermal feedback[34,35]. Once the current is reduced, the detector returns to the superconducting state. As illustrated in Fig. 1c, the voltage probes in the 4-terminal scheme are connected to a room-temperature low-noise amplifier, the output of which is fed to an oscilloscope or an analog-to-digital converter to measure the voltage over time induced by the photons. When the MATBG transitions into the resistive state upon photon absorption, we register a spike in $V_{ph}$, the detector resets itself, and we can measure the statistics of counts as a function of bias voltage, laser power and temperature.

To perform the photoresponse measurements, we mounted the MATBG device in a dilution refrigerator and provided optical excitation with a 1550-nm laser diode. The beam was collimated in the sample space (~ 4 mm spot-diameter) allowing illumination of the entire device area. The incident laser power was then controlled using a programmable optical attenuator (see Methods). In our experimental setup both bias and readout leads were heavily filtered in order to ensure millikelvin electron temperature at the sample stage. The constrained electrical bandwidth available in the experiment imposes limitations on the maximum detector count rate and the speed of the reset circuitry but still allows to properly study the statistics of the photo-induced counts. We extensively describe the optoelectronic setup employed in our experiment and the method used to register the counts in section C and D of the Supplementary Information.

Fig. 2a illustrates an example of the photovoltage traces $V_{ph}(t)$, measured over time across the MATBG detector when it is exposed to the laser beam radiation in the configuration described in Fig. 1c. We observe voltage spikes, emerging as we increase the incident laser power, which we attribute to photo-induced switching events from the superconducting to the normal state. To confirm the origin of the voltage spikes which are the 'clicks' of our detector, we investigate their stochastic nature by producing histograms of counts with 1-s bins and extracting the mean ($\mu_{hist}$) and variance ($\sigma^2_{hist}$) of the sampling distribution[36]. As reported in the inset of Fig. 2b, for all the histograms the mean equals the variance, as prescribed by a Poisson process. We further demonstrate the agreement with this statistic by plotting on top of the histograms the Poisson distribution with the extracted $\mu_{hist}$ and $\sigma^2_{hist}$ (solid lines). The excellent agreement between the experimentally registered counts and the statistical model confirms that our observation is indeed compatible with the photon shot noise generated by the highly attenuated CW laser source.

Additionally, we examine the average click height as a function of the bias voltage, $V_{bias}$. These results are overlaid on the *I-V* curve (top inset in Fig. 2a) which was measured in the configuration described in Fig. 1c. We find that the generated photovoltage closely matches the voltage in the normal state across all explored bias voltages: $V_{ph}(V_{bias}) \approx V(V_{bias})$. This observation indicates that the incident photons induce a complete transition of the MATBG detector from its superconducting state to the normal state.

**Single-photon sensitivity by superconducting MATBG**
To investigate the observed photoresponse, in Fig. 3a we compare the photon count rate PCR (counts recorded per second) for different bias voltages ($V_{bias}$) without light (empty dots) and with an excitation wavelength of $\lambda = 1550$ nm for different laser powers (filled dots). When the detector operates at a bias voltage far from the transition ($V_c$), the PCR is orders of magnitude higher upon illumination than in the dark, while as $V_{bias}$ approaches $V_c$ a sudden increase in false-positive (dark) counts occurs, ultimately dominating the detector's response. We fit the PCR vs. $V_{bias}$ curve under illumination with a sigmoid function (solid line in Fig. 3a lower panel) and observe that the experimental data exhibit a tendency to saturation at $V_{bias} \sim$ 0.997 $V_c$. These saturations are intrinsic to the process of SPD, rather than extrinsic, to e.g. speed of the measurement circuitry (Supplementary Information, Section F), and resemble the photon counts in superconducting-nanowires single-photon detectors[7] (SNSPDs). In SNSPDs, the saturation of the PCR as a function of current bias indicates that the internal detector efficiency[37], without coupling, reaches unity. Conversely in our experiment, the PCR curve does not entirely saturate, implying that the intrinsic efficiency attained is not 100%.

We can demonstrate that the registered counts are triggered by single near-infrared photons. For this purpose, we explore how the count rate evolves as a function of the CW laser power over several orders of magnitude at two different bias points. To provide a quantitative description of the light-induced count rate we estimate the power density incident on the MATBG ($P_L$) in the approximation of a gaussian beam[38]. From $P_L$ we can calibrate the number of incident photons per μm² in a time window $\tau$ as: $\langle N_{photon} \rangle = \tau \cdot P_L / h\nu$, where $h\nu = 1.28 \cdot 10^{-19}$ $J$ is the energy of a single photon at $\lambda = 1550$ nm. Choosing $\tau = 5$ ms which is close to the typical detector recovery time, a laser power density of $P_L = 10$ aW/μm² corresponds to $\langle N_{photon} \rangle = 0.4$ photons incident per μm² in a time window of 5 ms (see Supplementary Information, section C). Under illumination with a weak coherent light source[36], the probability of detecting $m$ photons in a detection time window reduces to $\sim \langle N_{photon} \rangle^m / m!$. Fig. 3b shows the detection probability in a time window $\tau$ (PCR· $\tau$) as a function of $\langle N_{photon} \rangle$. The measured detection probability increases linearly with $\langle N_{photon} \rangle$ over > 3 orders of magnitude with an offset due to the dark counts, demonstrating single-photon sensitivity of the MATBG superconducting detector[1]. As reported for other SPDs[39], we observe that the count rate deviates from linearity at low photon fluxes, when it enters the noise level defined by the dark counts and at high photon fluxes when it saturates due to the limited bandwidth of measurement circuitry.

Importantly, the trace measured at 0.995 $V_c$ shows the same overall behavior as the one measured at 0.989 $V_c$ with higher detection probability and dark count rate due to the increase of the intrinsic quantum efficiency as we approach $V_c$. In the section D of the Supplementary Information we show several raw photovoltage time traces at different $V_{bias}$ and $P_L$ from which we extract the PCR reported in Fig. 3 and detail the method used to register the counts in the MATBG detector. For completeness, we also demonstrate single-photon sensitivity under pulsed light excitation (Supplementary information, Section E).

**Detector performance at higher temperatures**
To provide further insights on the photodetection mechanism in MATBG, in Fig. 4a we present the PCR versus $V_{bias}$ at six different temperatures ranging from 35 mK to 800 mK, both with and without laser excitation (filled and empty dots, respectively). The PCRs with illumination are consistent with the sigmoid function and exhibit a tendency to saturation for $V_{bias} \sim 0.997$ $V_c$. Using the linear scaling of the PCR with laser power (Supplementary Figure S14), we confirm the SPD from our MATBG device up to $\sim 0.7$ K. The single-photon PCR eventually vanishes when temperature rises up to 0.8 K, at which the dark count dominates the PCR. Fig. 4b plots the SPD efficiency as a function of $V_{bias}$ and the dark count rate (PCR without illumination) at various temperatures. Here the efficiency is defined as the ratio of counts detected per second to photons incident per second in the area ($A \sim 16$ μm²) between the two voltage probes (white dashed box in the optical image of Fig. 1f). Interestingly, the dark count rate (right-hand-side of the y-axis) exhibits two distinct $V_{bias}$ dependence above and below $V_{bias}$ = 0.998 $V_c$. When $V_{bias}$ > 0.998 $V_c$, a sharp increase in dark counts occurs. This justifies the abrupt rise of PCR under illumination when $V_{bias}$ ~ $V_c$. When $V_{bias}$ < 0.998 $V_c$, the dark counts rise gradually due to background photons coupling through the optical fiber connected at the room-temperature optical port[40]. Notably, the detection efficiency on the left-hand-side of the y-axis is at maximum and gradually decreases as the temperature rises, akin to observations in other SPDs[37,40]. To further investigate this trend, we extract the efficiency at three different $V_{bias}$ from the sigmoid fit of Fig. 4a, and plot them against temperature in Fig. 4c. The efficiency decreases as temperature rises. We attribute this to the increase of the thermal conductance. Elevated temperatures enhance heat transfer out of the electrons[41] reducing the probability of latching into the resistive state by a self-heating effect. This argument is supported by thermal transport measurements in the superconducting state on the same system, which report a rapid increase in thermal conductivity within the range 35 mK < $T$ < 800 mK (ref.[30]). Fig. 4d plots the trade-off between SPD efficiency and dark count rate at various temperatures to determine the optimal operating condition of our MATBG detector. We observe that the most favorable SPD performance is achieved in the plateau region, where the efficiency approaches its maximum value while maintaining a low dark count rate[11].

**Discussion**

In addition to the demonstration of SPD, our experiment offers insight to the MATBG superconductivity via its interaction with photons. Since the incident photon energy (~ 0.8 eV) significantly exceeds the flat bands' width[42] (~ 10 meV) and the superconducting gap's size[43,30] (~ 1 meV), we can approximate MATBG's absorption to be the same as bilayer graphene (~ 4.6%). With the measured PCR at the saturation plateau in Fig. 4b, we estimate the internal efficiency of our SPD as ~ $10^{-3}$/0.046 ~ 0.022. Two factors can limit the internal efficiency in our setup. Firstly, the effective area of the MATBG contributing to the photoresponse could be much smaller than the entire area of the device. This argument would agree with the measurements of twist angle inhomogeneity by local probes techniques on MATBG[44,45]. Twisted moiré materials are characterized by an intrinsic disorder due to the local variation of twist angle within the same sample. Due to the relaxation of the lattice structure, the atoms between the two twisted layers of graphene may not align to the same global twist angle, resulting in a narrower superconducting area in MATBG. If we assume an internal efficiency of ~ 1 at the saturation plateau, we can estimate a lower limit for the effective superconducting area contributing to the photoresponse as $A_{eff}$ ~ $0.022A \approx 0.35$ μm$^2$. If the superconducting channel fully percolates between the two voltage probes, spaced about 3 μm apart, we estimate the channel's width to be approximately 120 nm. This scenario would be similar to the SNSPDs in which the absorbed photon generates a resistive domain of quasiparticles to produce a readout signal[6]. Unlike SNSPDs, however, the hotspot in MATBG would expand with self-sustained Joule heating, leading to a complete breakdown of superconductivity across the whole superconducting path, as observed in the inset of Fig. 2a. Hence, heat dissipation, which reduces the latching probability, could be the second factor limiting the internal efficiency, as supported by the temperature-dependence data in Fig. 4c. Future applications may explore SPD readout mechanisms that do not involve a complete switching of the entire MATBG device into the normal state and could employ more sensitive probes to monitor the light-induced changes in voltage or kinetic inductance[10].

In conclusion, our experimental work has successfully demonstrated that the superconducting state of MATBG can be used to detect single near-infrared photons. This result strongly motivates further investigation to extend single-photon capability to lower energies using MATBG and other low-carrier density graphene-based superconductors[24–27]. Pursuing this route necessitates further research effort to understand the intricate interplay between incident photons and these novel superconducting phases. Our investigation has contributed significant insights into the physical process underlying the observed MATBG's photoresponse. These insights will play a pivotal role in the development of theoretical models and in the design of innovative quantum devices that exploit the unique characteristics of these materials, ultimately advancing the field of quantum technology.

Acknowledgements:

We acknowledge fruitful discussion with Paul Seifert and Robert H. Hadfield. D.K.E. acknowledges funding from the European Research Council (ERC) under the European Union's Horizon 2020 research and innovation program (grant agreement No. 852927), the funding from the EU EIC Pathfinder program, project FLATS (grant agreement No. 101099139) and the German Research Foundation (DFG) under the priority program SPP2244 (project No. 535146365). K.W. and T.T. acknowledge support from the Elemental Strategy Initiative conducted by the MEXT, Japan (Grant Number JPMXP0112101001) and JSPS KAKENHI (Grant Numbers 19H05790, 20H00354 and 21H05233).

Author contributions:

**Supplementary Information** is available for this paper.

**Correspondence and requests for materials** should be addressed to D.K.E.

**Competing interest.** Authors claim no competing interest.

**Data availability.** The data that support the findings of this study are available from the corresponding author upon reasonable request.

**Code availability.** The code that supports the findings of this study is available from the corresponding author upon request.

**Methods**

Device fabrication
The MATBG devices were fabricated using the "cut-and-stack" technique. A stamp made of propylene carbonate (PC) and polymethyl siloxane (PDMS) was prepared and mounted on a glass slide. The stamp was used to pick up the top layer of hexagonal boron nitride (hBN). The hBN layer was then used to pick up the two halves of graphene, which had been pre-cut using an atomic force microscopy (AFM) tip. The graphene halves were carefully rotated to a target twist angle of 1.1°. To complete the heterostructure, the entire stack was fully encapsulated with a bottom hBN layer. A graphite layer was added at the bottom of the stack, serving as a local back-gate for the device. The full stack was then deposited onto a Si/SiO2 chip and etched into a Hall bar geometry. Finally, edge contacts made of Cr/Au (5/50 nm) were evaporated onto the device to establish electrical connections.

Photoresponse measurements.
To perform the photoresponse measurements, we placed the device on the cold finger of a dilution refrigerator (BlueFors-SD250), housed in a gold-coated oxygen-free copper box. The dilution refrigerator (base temperature of 35 mK) was optimized for low-frequency transport of low-carrier density two-dimensional superconducting materials. A two-stage RC low-pass filter was mounted at the 1-K still plate of the dilution refrigerator combined with an additional RF filter mounted at the mixing chamber stage to ensure millikelvin electron temperature and reject high frequency noise (Supplementary Figure S6). The overall bandwidth of the readout was < 1 kHz (Supplementary Figure S8). In the SPD experiment described in Fig. 1c, the bias voltage was applied at the source contact with a voltage generator (Keithley 2400) in series with a 1/1000 voltage divider ($R_1$ = 1MΩ, $R_2$ = 1kΩ for device A). The voltage probes were connected to a room-temperature 1-MHz-bandwidth low-noise amplifier (SR-560). A room-temperature low-pass filter with a sharp cut-off at ∼ 1-10 kHz was used for rejecting white noise outside of the readout bandwidth. The amplified signal was fed to a sampling oscilloscope with variable bandwidth up to 600 MHz (UHF-Scope Zurich Instrument) or a 100-kHz-bandwidth analog-to-digital converter (UHF-Aux In Zurich Instrument). The optical excitation was provided by a 1550-nm laser diode (Taiko PDL M1). The light was fed into the dilution refrigerator to the MATBG detector, through a single-mode optical fiber coupled with a collimator mounted few centimeters on top of the sample space, allowing illumination of the entire device area ∼ 4 mm spot-diameter (Supplementary Figure S7). To control the incident laser power, a programmable optical attenuator (JGR OA5 l) was employed, enabling precise adjustment over several orders of magnitude. Extensive description and schematics of the optoelectronic setup employed in the experiment are available in the section C of the Supplementary Information.

Transport measurements.
The longitudinal resistance $R_{xx}$ was measured using standard low-frequency lock-in techniques (Stanford Research SR860). To control the carrier density, a voltage was applied to the graphite metallic gate using a Keithley 2400 voltage source connected in series with a 100 MΩ resistor. For measuring the current-biased *I-V* curves in the 4-terminal configuration, the bias current was supplied at the source contact using a voltage generator (Keithley 2400) connected in series with a bias resistor of 10 MΩ. The voltage probes were connected to a room-temperature low-noise preamplifier (SR-560), and the output signal was measured using a digital multimeter (Keithley 2700).

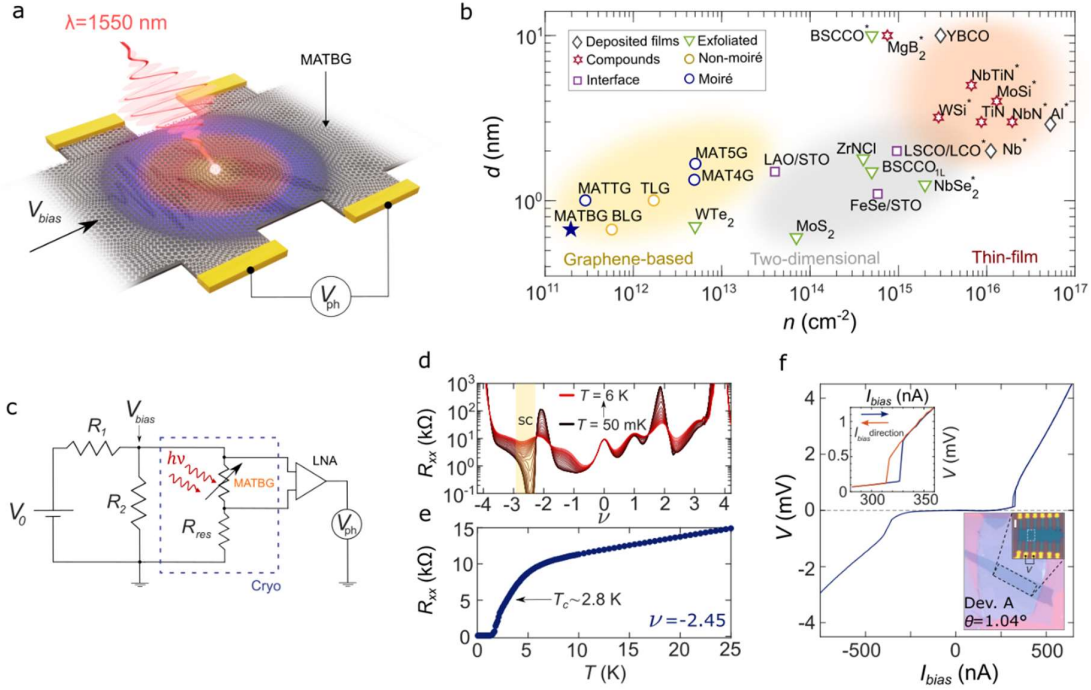

Fig. 1. **Superconducting MATBG as an ultra-sensitive material for SPD.** (a) The near-infrared photon, incident on the voltage-biased MATBG device, breaks Cooper pairs and generates a photovoltage output, $V_{ph}$. (b) Logarithmic plot of film thickness $d$ versus carrier density $n$ for various superconductors. Data are taken from refs.[28,30,24,26,27,25,19,46–50]. The plot includes deposited thin film superconductors (Nb, Al, YBa$_2$Cu$_3$O$_{7-\delta}$), compound thin film superconductors (NbTiN, MoSi, WSi, TiN, NbN), interfacial 2D superconductors (LaAlO$_3$/SrTiO$_3$, FeSe/SrTiO$_3$, La$_{1.55}$Sr$_{0.45}$CuO$_4$/La$_2$CuO$_4$), exfoliated 2D superconductors (ZrNCl, Bi$_2$Sr$_2$CaCu$_2$O$_{8+\delta}$, NbSe$_2$, MoS$_2$, WTe$_2$) as well as moiré (MATBG, magic-angle twisted trilayer graphene, magic-angle twisted four-layer graphene, magic-angle twisted five-layer graphene) and non-moiré (Bernal bilayer graphene, rhombohedral trilayer graphene) graphene-based superconductors. The red, gray and yellow shaded regions serve as rough distinction between thin-film superconductors, two-dimensional superconductors and graphene-based superconductors, respectively. The asterisk indicates the superconducting materials which have been previously used for photodetection applications[3,6,18–20]. In the case of Bi$_2$Sr$_2$CaCu$_2$O$_{8+\delta}$ SPD was demonstrated but not in the monolayer limit[19]. (c) Simplified circuit diagram used to measure the photoresponse of the MATBG detector. The voltage divider ($R_1$ = 1MΩ, $R_2$ = 1kΩ for device A) provides a voltage bias ($V_{bias}$) across the source and drain contacts of the device. The incident photons induce voltage spikes in the MATBG detector (sketched as a variable resistor) which are recorded using an oscilloscope or an analog-to-digital converter. $R_{res}$ ~ 53 kΩ is the residual resistance (arising from the contact resistance and the metallic leads). (d) Longitudinal resistance $R_{xx}$ of device A ($\theta$ = 1.04°) as a function of the filling factor $\nu$ for successive temperatures $T$ ranging from 50 mK to 6 K. A pronounced superconducting state is observed for -3 < $\nu$ < -2. (e) $R_{xx}$ vs. $T$ at the optimal doping of $\nu$ = -2.45. (f) I-V curve measured at the optimal doping, displaying a hysteretic behavior with respect to the sweeping direction of the bias current, highlighted in the top inset. In the bottom inset, the optical image of the MATBG device. The area measured by the two voltage probes is marked by the white dashed box $A$ ~ 16 μm$^2$. Scale bar: 3 μm.

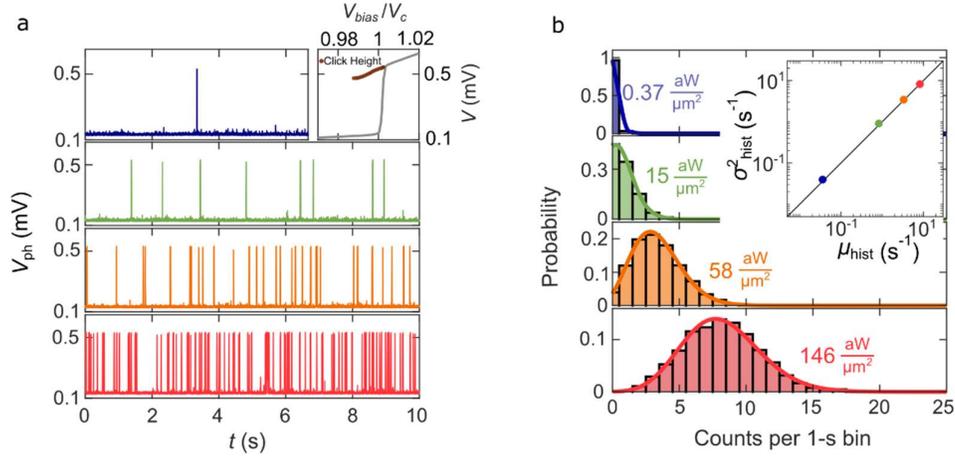

Fig. 2. **Statistics of the light-induced 'clicks'.** (a) Raw photovoltage time traces, $V_{ph}(t)$ measured at increasing laser powers for $\lambda = 1550$ nm. Top right inset: Average click height measured as a function of $V_{bias}$. The click heights are overlaid on the *I-V* curve, measured in the configuration described in Fig. 1c. (b) Histograms of counts in 1-s bins for the same laser powers in (a) measured over $\sim 10^3$-s time window. The inset shows that the extracted variance of counts $\sigma^2_{hist}$, equals to its mean $\mu_{hist}$. The agreement with the Poisson statistic is confirmed by plotting on top of the histograms the Poisson distribution with the extracted $\mu_{hist}$ and $\sigma^2_{hist}$ (solid lines).

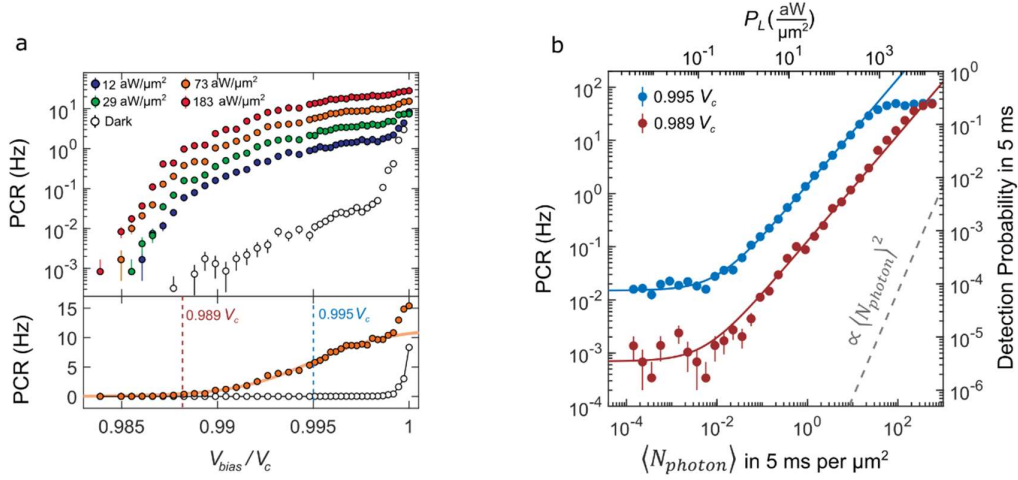

Fig. 3. **Single-photon sensitivity by superconducting MATBG.** **(a)** Top panel: Photon count rate, PCR as a function of voltage bias $V_{bias}$ for four different laser powers (filled dots) and in the dark (empty dots). Bottom panel: PCR vs. $V_{bias}$ for $P_L$ = 73 aW/μm² on a linear scale. The orange line is a fit with a sigmoid function. The PCR shows a tendency to saturation at ~ 0.997 $V_c$. The vertical dashed lines are the bias points at which we performed the PCR vs. $P_L$ measurements reported in Fig. 3b. **(b)** PCR versus the average incident photon number $\langle N_{photon}\rangle$ in a 5-ms time window per μm² for two different bias points ($V_{bias}$ = 0.995 $V_c$ and $V_{bias}$ = 0.989 $V_c$). On the top x-axis the corresponding incident CW power density $P_L = \langle N_{photon}\rangle \cdot h\nu/\tau$ and on the right y-axis the corresponding detection probability in a 5-ms time window (PCR·$\tau$). The solid lines are linear fits (with an offset due to dark counts), showing that the detection probability evolves linearly with $\langle N_{photon}\rangle$. The gray dashed line depicts a quadratic power dependence.

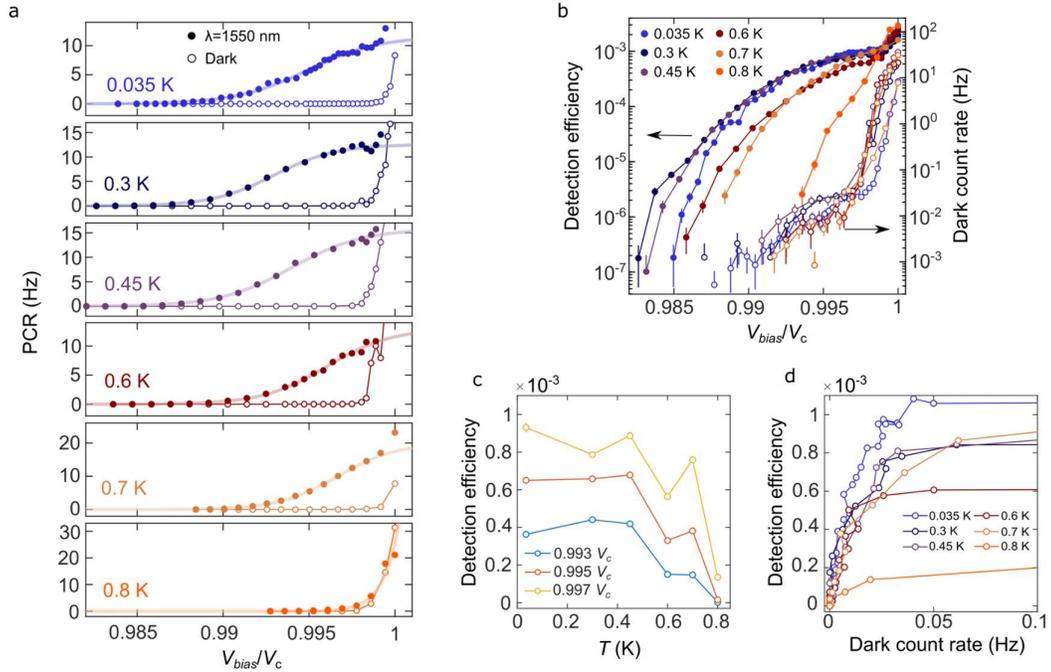

Fig. 4. **Detector performance at higher temperatures. (a)** Photon count rate, PCR as a function of voltage bias $V_{bias}$ and temperature $T$ upon illumination (filled dots) and in the dark (empty dots). The continuous lines are fit with the sigmoid function. **(b)** Filled markers: detection efficiency vs. $V_{bias}$ at different temperatures. Empty markers: dark count rate vs. $V_{bias}$ at different temperatures. The detection efficiency is defined as the ratio of counts detected per second to photons incident per second in the area ($A \sim 16$ μm$^2$) between the two voltage probes. **(c)** Detection efficiency vs. $T$ for three different bias points extracted from the sigmoidal fit of Fig. 4a. **(d)** Trade-off between detection efficiency and dark count rate for different temperatures.

# Supplementary information: 'Ultra-low carrier density superconducting bolometers with single photon sensitivity based on magic-angle twisted bilayer graphene'

G. Di Battista[1], K.C. Fong[2,3], A. Díez-Carlón[1], K. Watanabe[3], T. Taniguchi[4] and D. K. Efetov[1,5,6]*

1. Fakultät für Physik, Ludwig-Maximilians-Universität, Schellingstrasse 4, München 80799, Germany
2. Department of Physics, Harvard University, Cambridge, MA 02138, USA
3. Quantum Engineering and Computing Group, Raytheon BBN Technologies, Cambridge, MA 02138, USA
4. Research Center for Functional Materials, National Institute for Materials Science, 1-1 Namiki, Tsukuba 305-0044, Japan
5. International Center for Materials Nanoarchitectonics, National Institute for Materials Science, 1-1 Namiki, Tsukuba 305-0044, Japan
6. Munich Center for Quantum Science and Technology (MCQST), München, Germany

*E-mail : dmitri.efetov@lmu.de

## Table of Contents

## A. Extended transport characterization for device A

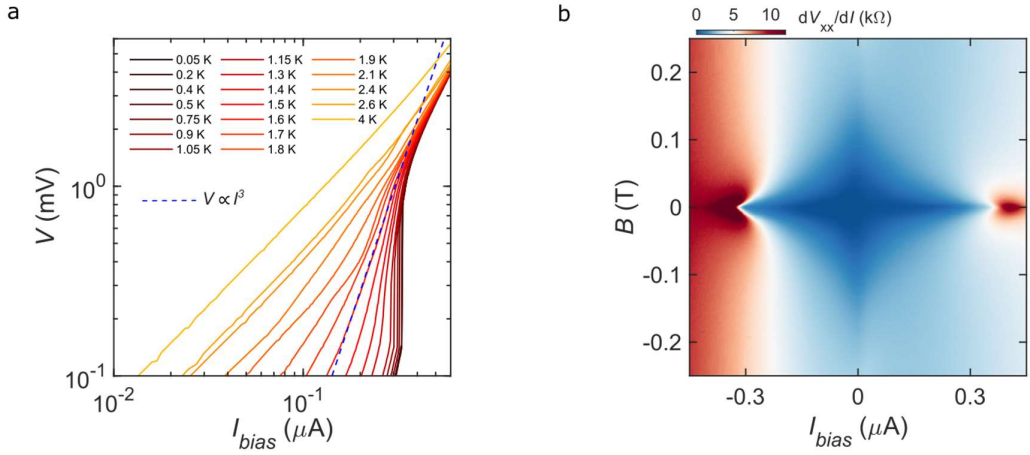

Supplementary Figure S1.| **Full characterization of the superconducting state of device A at *v* = -2.45. (a)**. Current-voltage (*I-V*) curves measured at different temperatures and plotted in logarithmic scale. The logarithmic scale helps in determining the Berezinskii-Kosterlitz-Thouless transition temperature ($T_{BKT}$ ~ 1.6 K) by fitting the data to a power law $V \propto I^3$ (represented by the blue dashed line). **(b)** Differential resistance $dV_{xx}/dI$ as a function of bias current $I_{bias}$ and out-of-plane magnetic field *B* for *v*=-2.45. The ac excitation current used for this measurement is $I_{ac}$ = 2 nA. The superconducting state is completely smeared out by magnetic field at *B* ~ 150 mT.

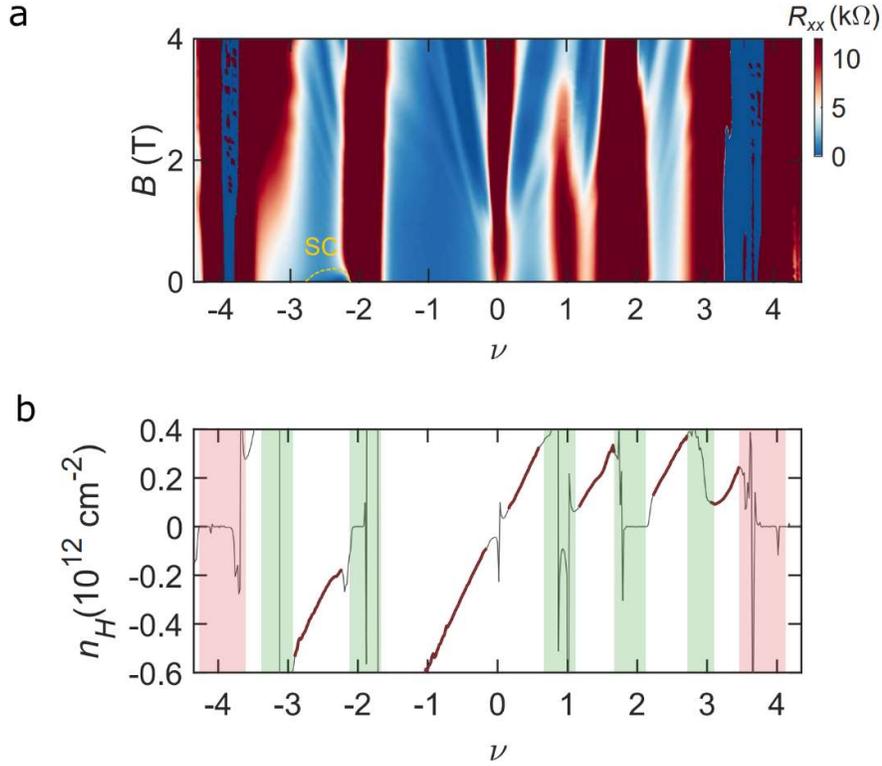

Supplementary Figure S2.| **Transport characterization with out-of-plane magnetic field for device A. (a)** Landau fan diagram measured at 35 mK with an excitation current of 10 nA and an applied out-of-plane magnetic field $B$ ranging from 0 to 4 T. **(b)** Low-field Hall density measurements. The light-gray line trace shows the Hall carrier density $n_H$ versus the moiré band filling factor $\nu$ measured at 500 mT. The light green stripes indicate the position of correlated states while the light red ones the band insulating states. The thick red lines delineate the regions where the Hall carrier density exhibits a linear relationship with the filling factor. Notably, at the integer fillings corresponding to the correlated states, we observe resets of the Hall carrier density, resulting in the presence of extremely low carriers involved in the conduction process. Specifically, for the doping used in the single-photon detection experiments ($\nu$ between -2 and -3) we expect a carrier density $n_H \sim 10^{11} \text{cm}^{-2}$.

## B. Hysteretic *I-V* curves in superconducting MATBG devices

For this project we have produced several MATBG superconducting devices with the 'cut-and-stack' procedure described in the methods section. Among all the superconducting MATBG we have selected 3 devices which featured sharp superconducting transitions with hysteretic *I-V* characteristics and investigated their photoresponse.

Device A and C have a single bottom graphite back gate, while device B has an additional graphite top-gate which was picked up at the first step of the stacking process. The global twist angles measured from transport data for device A, B and C are $\theta = 1.04°$, $\theta = 1.03°$ and $\theta = 1.16°$ respectively.

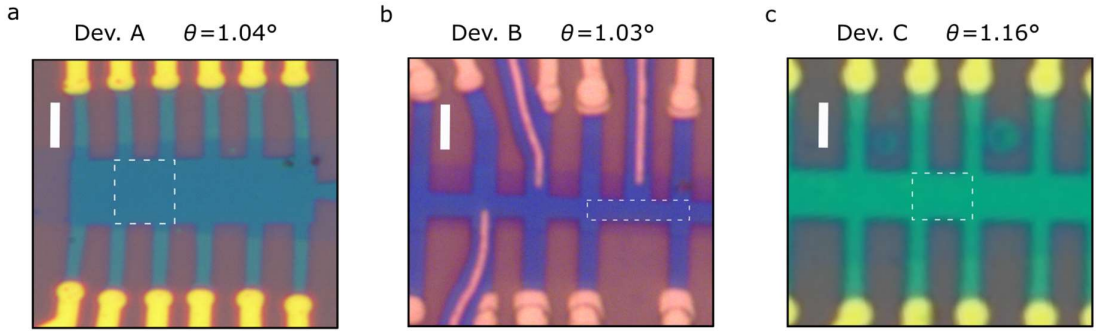

Supplementary Figure S3.| **Optical images of the measured devices.** The scale bar in all the images is 3 μm. Device A and C have a single bottom graphite back-gate while device B has a double graphite gate.

As detailed in the main text, we attribute the presence of a hysteretic loop in the *I-V* curves to a current-induced self-heating hotspot when the MATBG is in the normal state. In Supplementary Figure S4 we show the *I-V* curves of devices A, B and C at different temperatures and gate voltages (corresponding to different carrier densities). In all devices the hysteresis loop disappears for temperatures $T \sim 1$ K. We also point out that the applied gate voltage provides high tunability on the critical current and on the width of the hysteretic loop. This represents an important tuning knob to engineer and design the electronic circuit.

In order to control the hysteretic loop in the MATBG device, we implement the voltage-biased scheme detailed in Fig. 1c of the main text. In the voltage-biased scheme, the increase of resistance induced when the MATBG detector is in the normal state, reduces the current flowing in it and brings the device back to the superconducting state. In Supplementary Figure S5 we observe the *I-V* curves for device A, B and C measured in the current-biased scheme (top) and in the voltage-biased scheme (bottom). We notice that in all devices the hysteresis loop present in the current-biased scheme is completely closed in the voltage-biased one. The simple reset circuitry described here, prevents permanent 'latching' of the detector in the normal state and allows us to reset it after photo-absorption.

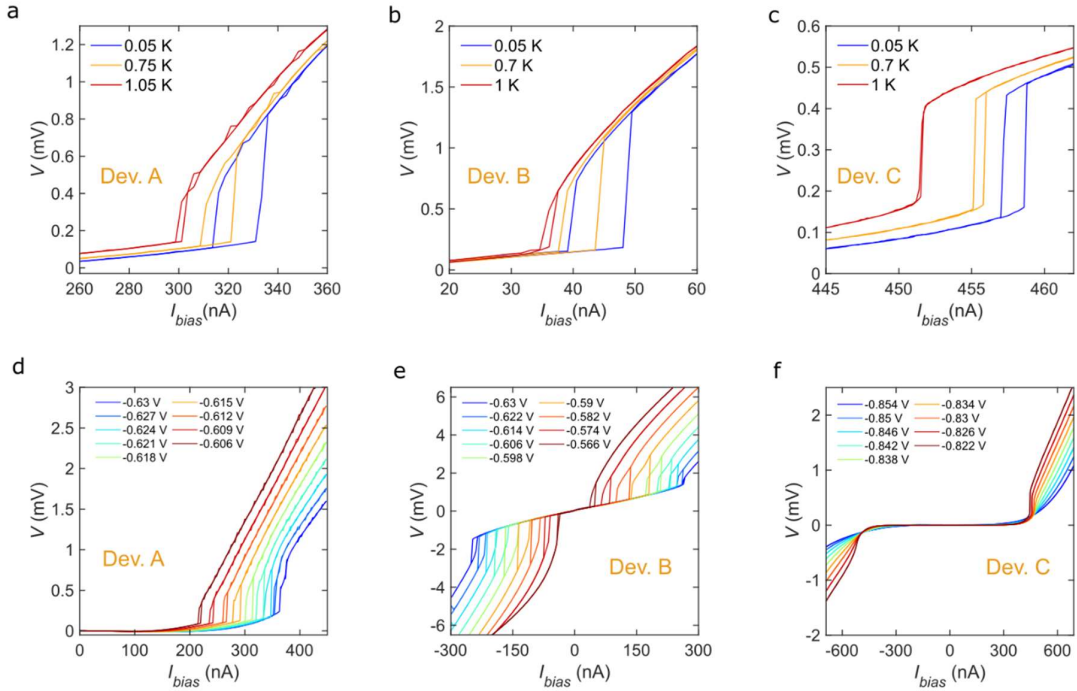

Supplementary Figure S4.| **Current-biased *I-V* curves at different temperatures and gate voltages.** **(a)-(c)** Current-biased *I-V* curves at 3 different temperatures for device A, B and C respectively. The *I-V* curves are measured at the doping (gate voltages) used for photodetection which are -0.620 V, -0.566V and -0.8257 V for device A, B and C respectively. In all the 3 devices the hysteresis loop disappears for temperatures ~ 1 K. **(d)-(f)** Current-biased *I-V* curves at different gate voltages (carrier densities) and $T$ = 35 mK within the superconducting dome for device A, B and C respectively.

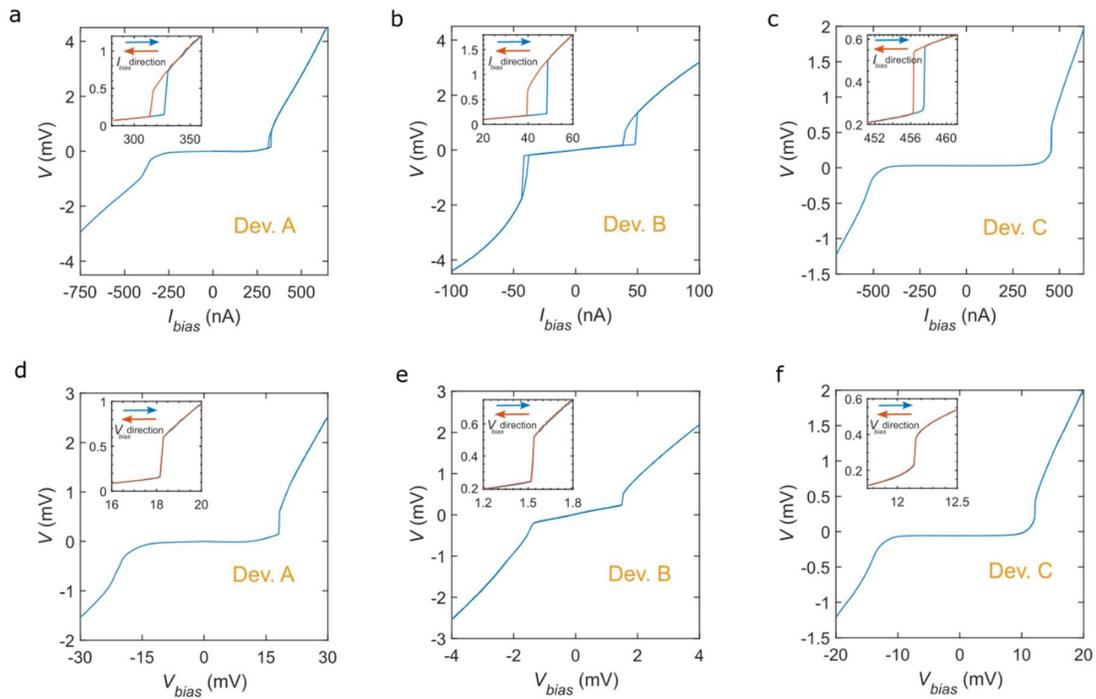

Supplementary Figure S5.| **Current-biased and voltage-biased *I-V* curves. (a)-(c)** Current-biased *I-V* curves for device A, B and C. The bias current is provided by a voltage source in series with a 10MΩ resistor. The *I-V* curves are measured at the doping used for photodetection which are -0.620 V, -0.566V and -0.8257 V for device A, B and C respectively. **(d)-(f)** Voltage-biased *I-V* curves for device A, B and C measured at the same doping. The bias voltage is provided by a voltage source in series with a 1/1000 voltage divider. As described in the main text the load resistor is much smaller than the residual resistance arising from the contact resistance and the metallic leads ($R_2 \ll R_{res.}$). $R_1 = 1$ MΩ, $R_2 = 1$ kΩ for device A and B while $R_1 = 100$ kΩ, $R_2 = 100$ Ω for device C.

## C. Optoelectronic setup

As explained in the Methods section an illustrated in Supplementary Figure S6, to perform the photoresponse measurements we placed the device in a dilution refrigerator and provided optical excitation with a 1550-nm laser diode coupled through a telecom single-mode optical fiber.

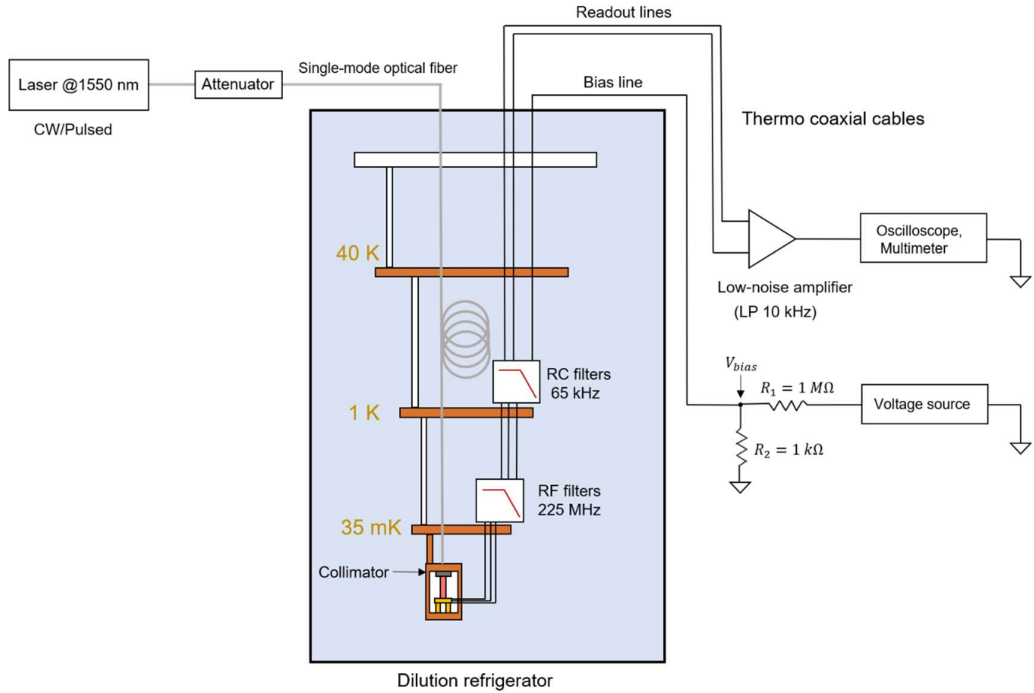

Supplementary Figure S6.| **Optoelectronic setup.** Schematics of the optoelectronic setup employed to measure the photoresponse in the MATBG superconducting detector.

# Beam profile at the sample stage

As shown in Supplementary Figure S7, in our setup we couple a telecom laser which emits an output power $P_{out}$ with a single-mode optical fiber designed for 1550-nm transmission. The fiber is then connected to a laser beam coupler which provides a collimated output with beam radius $w_0 \sim 2$ mm and Rayleigh range $z_R = \frac{\pi w_0^2}{\lambda} \sim 8$ m. In order to quantitatively describe the light-induced count rate on the MATBG detector we consider a Gaussian beam profile. In this approximation, the intensity profile $I$ as a function of the distance from the beam center $r$ and distance away from the end of the coupler $z$ reads[1]:

$$I(r,z) = I_0 \left(\frac{w_0}{w(z)}\right)^2 e^{-2(r/w(z))^2} \tag{C.1}$$

Where $w(z)$ is the value of the radius at a distance $z$ from the fiber given[1] by $w(z) = w_0\sqrt{1 + (z/z_R)^2}$ and $I_0 = 2P_{out}/(\pi w_0^2)$ is the total irradiance coming out of the laser source imposing the Gaussian normalization condition. In Supplementary Figure S7c we simulate $w(z)$ up to 1 m of distance from the fiber coupler. Since in our experimental configuration the device is around $z_0 \sim 3$ cm away from the fiber coupler ($\frac{z_0}{z_R} = 0.0037 \ll 1$) we can consider the beam to be collimated and replace in (C.1) $w(z) \simeq w_0 = 2$ mm. Since we align the device to be roughly at the center of the beam (C.1) reads $I(r = 0, z_0) = \frac{2P_{out}}{\pi w_0^2}$.

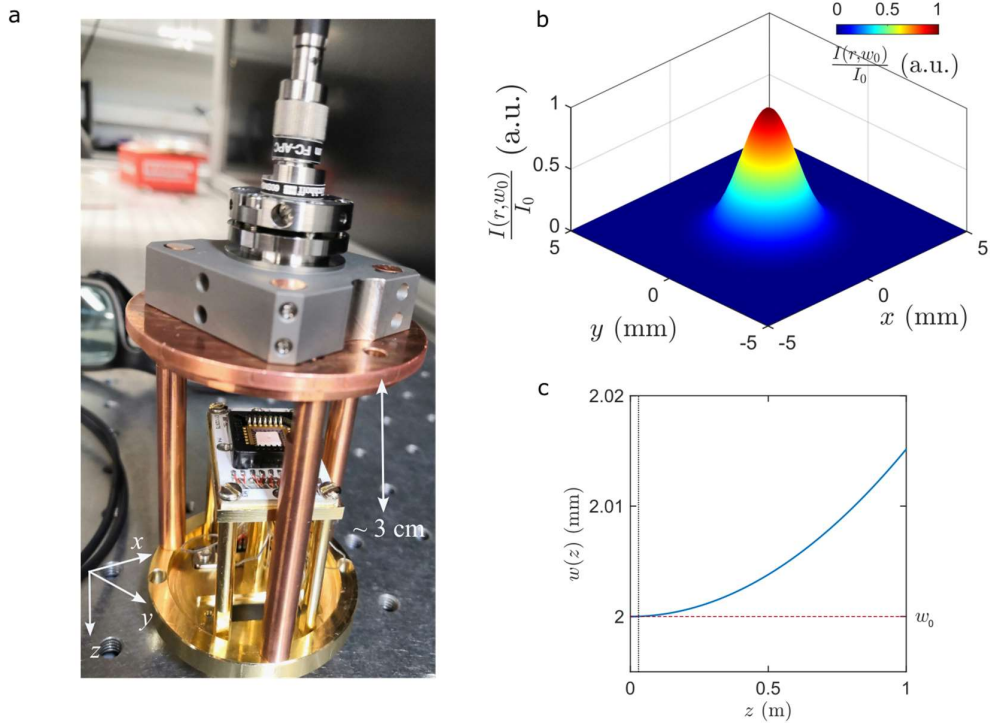

Supplementary Figure S7.| **Beam profile at the sample stage. (a)** Optical image of the experimental scheme. The single-mode optical fiber designed 1550-nm transmission is connected to a laser beam coupler which provides a collimated output. The sample is located at around 3 cm far from the fiber coupler. **(b)** 3D plot of the normalized beam intensity in the Gaussian beam approximation. **(c)** Simulation of the Gaussian beam radius $w(z)$ at a distance $z$ from the fiber coupler $w(z) = w_0\sqrt{1 + (z/z_R)^2}$. The black vertical line is the position of the sample $z_s = 3$ cm.

**Calculation of the power density incident on the MATBG device**

With the considerations made in the previous section, we can calculate the power density $P_L$ incident on the MATBG:

$$P_L = \frac{10^{-\frac{\eta}{10}} \cdot T_{fiber} \cdot P_{out}}{\frac{\pi}{2} w_0^2} \tag{C.2}$$

Where $P_{out}$ is the total power output coming out of the laser, $T_{fiber} = 0.021$ the effective transmission of the fiber and all the optical connections and $\eta$ the variable attenuation (in dB) we use to control the power incident on the device. In the single-photon measurements with the CW laser source we keep the laser power constant ($P_{out} = 11$ µW for device A) and scan $\eta$ between several order of magnitudes from 70 dB to 4 dB. For device A, a typical attenuation of 40 dB results in $P_L = 3.7 \frac{aW}{\mu m^2}$.

Given $P_L$, the average incident photon rate per unit time $\tau$ per µm² $\langle N_{photon} \rangle$ reads:

$$\langle N_{photon} \rangle = \tau \cdot \frac{P_L}{h\nu} \tag{C.3}$$

Where $h\nu = 1.28 \cdot 10^{-19}$ J is the energy of a single photon at $\lambda = 1550$ nm. For an attenuation of 40 dB and $\tau = 5$ ms, we expect $\langle N_{photo} \rangle = 0.14$.

In pulsed experiments, by changing the laser repetition rate ($f_{RR}$) we can control the number of photons carried on average by each pulse $\mu$ as:

$$\mu = \frac{1}{f_{RR}} \cdot \frac{P_L}{h\nu} \cdot \int_{-l_1/2}^{+l_1/2} dx \int_{-l_2/2}^{+l_2/2} dy\, e^{-2\frac{x^2+y^2}{w_0^2}} \tag{C.4}$$

Where $l_1 \sim 3$ µm and $l_2 \sim 5.3$ µm are the length and width of the area between the two voltage probes (white dashed box in the optical image of Fig. 1f). Since we assume the sample to be located at the center of the beam and $l_1, l_2 \ll w_0$ we can simplify (C.4) to:

$$\mu = \frac{1}{f_{RR}} \cdot \frac{P_L}{h\nu} \cdot l_1 l_2 \tag{C.5}$$

For a typical $P_{out} = 3$ nW, $f_{RR} = 100$ Hz, and attenuation of 13 dB, we obtain $\mu = 0.62$.

It is worth noticing that the calculated values are only an upper-bound estimation because the optical alignment is not controlled accurately in the cryogenic experiment. In particular it is possible that the sample is not perfectly located in the center of the beam and that the effective incident power is lower.

**Effective bandwidth of the electrical readout**

In this section we measure the overall bandwidth of the electronic readout available in our experiment which determines the distortion of the electrical pulses and limits the reset circuitry. Assuming that our electronic readout effectively behaves as an ideal RC low-pass circuit, we characterize its performance by determining the minimum rise time of electrical pulses and evaluating the available 3-dB bandwidth. To obtain these parameters under conditions closely resembling our single-photon detection experiment, we place a resistor (10 kΩ) into the sample space and monitor the voltage across the resistor using a 4-terminal configuration (Supplementary Figure S8b).

To measure the minimum rise time, defined as the length of time required for a signal to transition from the 10% to the 90% of the rising edge of the curve, we employ an arbitrary wavefunction generator (AWG) to generate square wave pulses with a frequency of 13.3 Hz. We record the voltage across the 10 kΩ resistor with an oscilloscope (Supplementary Figure S8a, b). By analyzing the square wave pulse (Supplementary Figure S8c) we extract the minimum rise time $t_r = 422$ μs. From this measurement, we calculate the effective 3-dB cut-off frequency as: $f_{3dB} = 0.35/t_r = 830$ Hz. Additionally, we directly measure the 3-dB bandwidth by applying a sinusoidal ac current to the resistor at various frequencies ranging from 3 Hz to 100 kHz using a lock-in amplifier. We measure the voltage generated across the 10 kΩ resistor with a lock-in amplifier in the same configuration (see Supplementary Figure S8d). From this measurement we can extract the 3-dB bandwidth defined as the frequency at which the initial amplitude drops by 3-dB or 0.707 of its initial value. We obtain $f_{3dB}$ = 738 Hz. The two different measurements give compatible results.

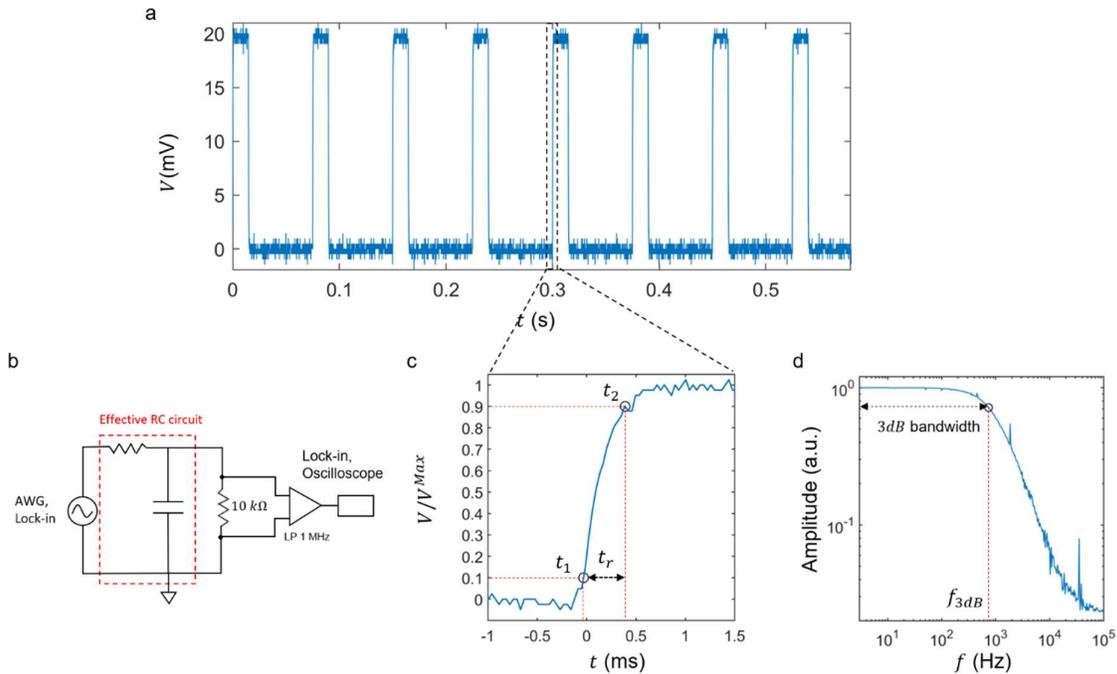

Supplementary Figure S8.| **Minimum rise time and frequency response magnitude of the electronical readout.** (a) Voltage signal measured with an oscilloscope across a 10 kΩ resistor when excited with a square wave of frequency 13.3 Hz. (b) Schematics of the circuit used to measure the rise time and the 3-dB bandwidth. (c) Zoom of the voltage traces from which we extract the rise time of the square wave pulse. (d) Measured frequency response magnitude of the readout.

In Supplementary Figure S9a we report an example of the photovoltage generation, $V_{ph}$, measured with an oscilloscope when the MATBG device is exposed to laser beam radiation of wavelength $\lambda$ = 1550 nm. The measurement is performed in the configuration described in Supplementary Figure S6 using a room-temperature low-pass filter with 10 kHz cut-off. Upon photon absorption, the MATBG detector transitions to the normal state, resulting in a maximum voltage output of $V_{ph}(V_{bias}) \approx V(V_{bias})$. Subsequently, the detector remains in the normal state for few ms (~ 1 ms for the trace in Supplementary Figure S9a) before the voltage bias circuit resets it to the superconducting state. It is worth noting that our observed pulse shape differs from the typical behavior observed in conventional superconducting single-photon detectors[2]. In those detectors, the generated photovoltage exhibits a rapid spike followed by a slower decay with a characteristic time ($\tau$) determined by the kinetic inductance of the superconducting circuit[3,4] or the intrinsic thermalization time[5,6].

We observe that the measured pulse rise time ($t_r$ = 356 μs) in our pulse shape (Supplementary Figure S9b) is extrinsically limited by the restricted bandwidth and compatible with the one measured with the 10 kΩ resistor. Similarly, the measured decay time is $t_d \sim t_r$. Despite the restricted bandwidth available in our experiment, we are still able to properly study the statistics of the photo-induced counts and demonstrate single-photon sensitivity by the MATBG detector.

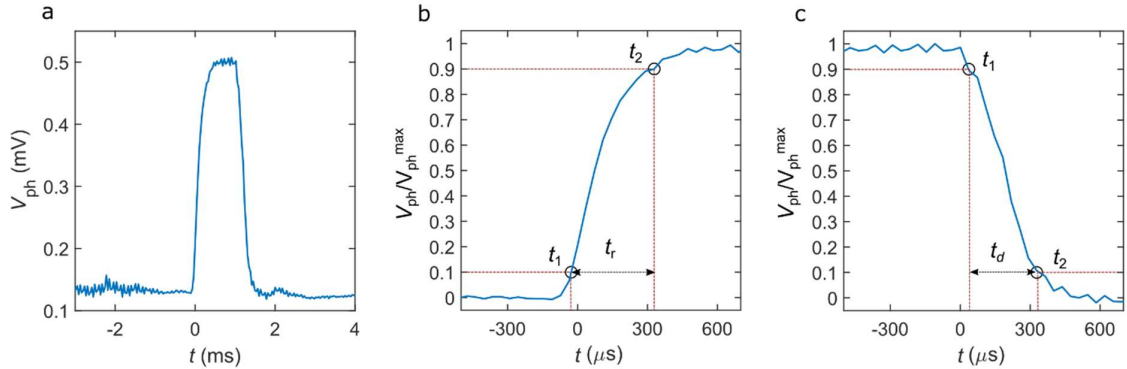

Supplementary Figure S9.| **Pulse shape, rise time and decay time for device A. (a)** Photovoltage pulse $V_{ph}$ measured in the MATBG photodetector at $V_{bias}/V_c \sim 0.989$ and $\lambda$ = 1550 nm with a single-shot oscilloscope. **(b)** Rise time $t_r$ = 356 μs measured from the pulse in **(a)** which results in an overall bandwidth of the electronic readout of < 1 kHz. **(c)** The decay time $t_d$ is similar to the rise time.

### D. Method of registering counts in the detector

In this section we show how we derive the plots shown in Fig. 3 from the raw data and detail the methods used to extract the counts of the MATBG detector. As described in the Methods section, we acquire photovoltage time traces with an analog-to-digital converter or an oscilloscope. From the raw traces (as the ones shown in Fig. 2a of the main text) we use a MATLAB script to count the number of detected events by setting a threshold ($V_{ph} > 0.4$ mV for device A) and a minimum distance between the clicks of 17 ms. We choose a minimum distance of 17 ms because the recovery time of the clicks varies as a function of the bias point from ~ 1 ms to ~ 17 ms. This limits the maximum measurable count rate to ~ 50 Hz. For the PCR vs $P_L$ measurements, the minimum distance between the clicks is 13 ms.

In Supplementary Figure S10a, we present the photon count rate (PCR) as a function of $V_{bias}$, measured at various laser powers (from no power up to 183 $\frac{aW}{\mu m^2}$) as in Fig .3a of the main text. In Supplementary Figure S10 (b)-(g) we also report the raw photovoltage time traces with and without illumination for six different bias points (vertical colored lines) from which we extracted the PCR vs. $V_{bias}$. In Supplementary Figure S11, we show the PCR vs. $P_L$ for $V_{bias}$ = 0.995 $V_c$ as in Fig. 3b of the main text and the raw photovoltage time traces for six different laser powers.

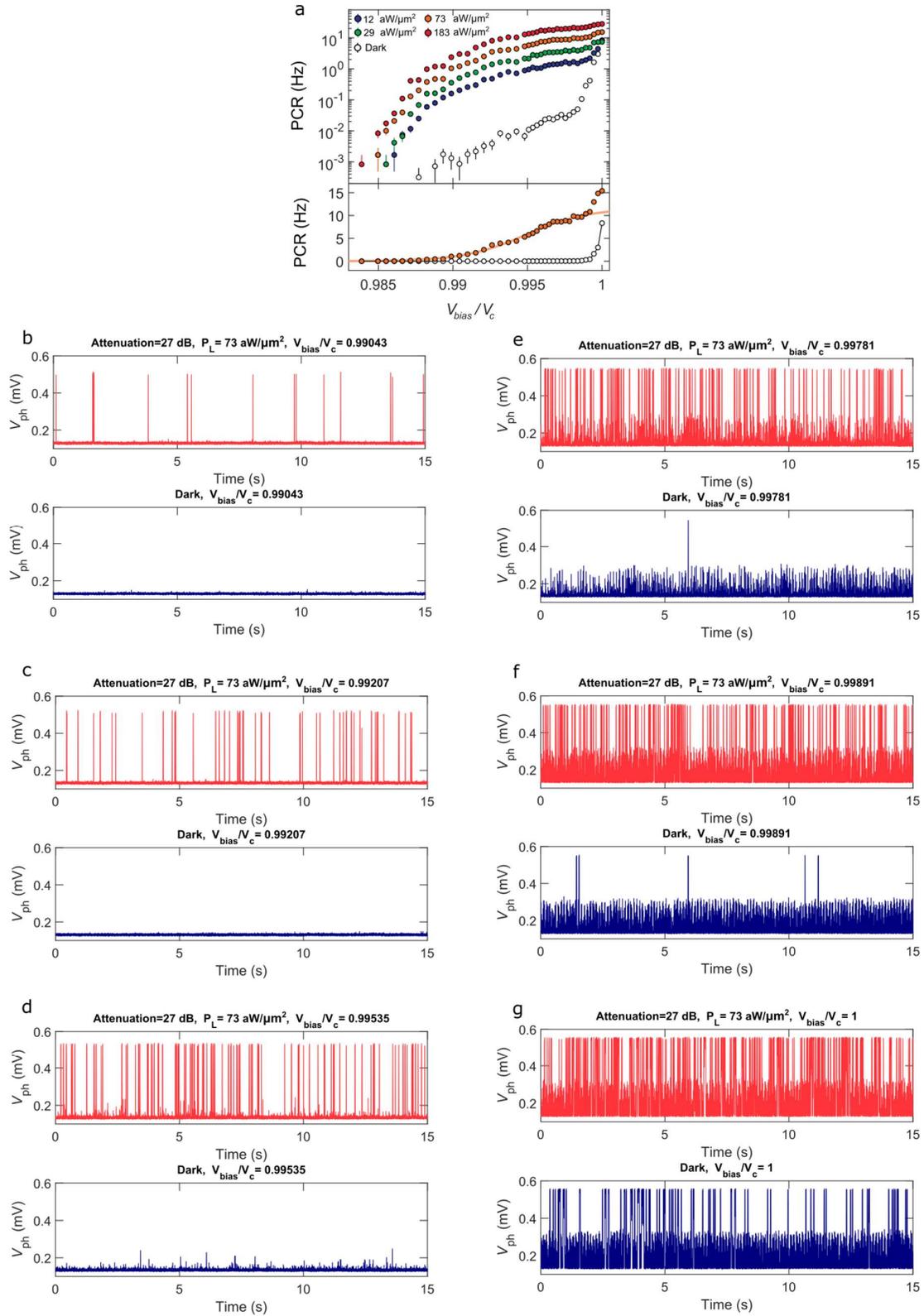

Supplementary Figure S10.| **Raw photovoltage time traces for different bias points. (a)** Photon count rate (PCR) vs. $V_{bias}$ measured for different laser powers as in Fig. 3a. **(b)-(g)** Raw photovoltage time traces with and without 1550 nm laser-illumination for different bias points.

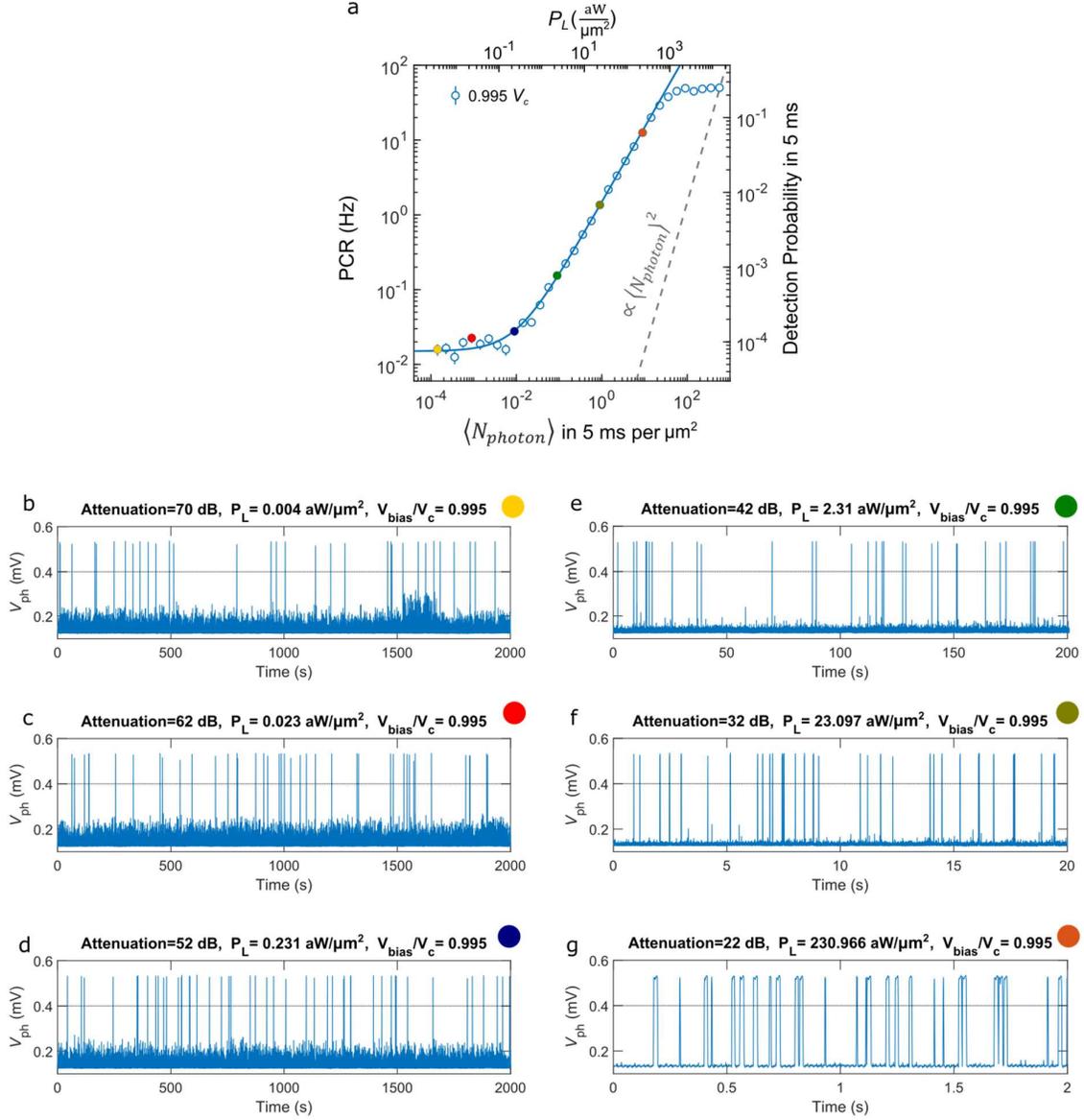

Supplementary Figure S11.| **Raw photovoltage time traces for different laser powers. (a)** Extracted photon count rate (PCR) versus laser power for $V_{bias} = 0.995\ V_c$ as shown in Fig. 3b of the main text. The colored dots are the selected laser powers for which we show the raw photovoltage time traces. **(b)-(g)** Raw photovoltage time traces measured over time for laser powers ranging over 5 orders of magnitude (laser attenuation from 70 dB to 22 dB). The black dashed line represents the threshold for counting the clicks. If the generated photovoltage is $V_{ph} > 0.4$ mV, we register a count in the detector. **(b)-(d)** Raw photovoltage time traces measured over 2000 second for attenuations of 70, 62 and 52 dB. In this range, the PCR (< 0.02 Hz) is not significantly affected by the increase of laser power, meaning that the observed counts are mostly due to false positive (dark) counts. **(e)-(g)**. Raw photovoltage time traces measured over 200, 20 and 2 seconds for attenuations of 42, 32 and 22 dB respectively. In these plots we scale the time duration of the traces inversely with the laser attenuation to facilitate the counting of the 'clicks' and allow to derive the PCR by eye. In this range, we observe that the PCR scales linearly with the incident power: 42 dB (PCR ~ 0.15 Hz), 32 dB (PCR ~ 1.5 Hz) 22 dB (PCR ~ 14 Hz).

## E. Single-photon sensitivity with pulsed light excitation

An independent way to cross-check the single-photon sensitivity observed under CW illumination is provided by measurements with pulsed light excitation. For this purpose, we use a ~ 50 ps laser source at $\lambda = 1550$ nm with broadly tunable repetition rate, $f_{RR}$. The pulsed laser source allows independent control of both the number of photons carried on average by each pulse ($\mu$) and of the frequency at which the pulses impinge on the device (inset of Supplementary Figure S12a). Having at disposal these tuning knobs, we demonstrate that the MATBG detector responds to pulses with less than one photon on average and confirm the linearity of single-photon sensitivity even under pulsed light excitation.

First, we measure the count rate at different $f_{RR}$ spanning over several orders of magnitude from 10 Hz to 1 MHz while fixing the number of photons carried on average by each pulse, $\mu$. Specifically, we keep $\mu = P_L A/h\nu\, f_{RR} = 0.62 < 1$ fixed by tuning simultaneously $f_{RR}$ and $P_L$. Here $A \sim 16$ $\mu m^2$ is the area between the two voltage probes. Arguably, the effective area contributing to the photoresponse is smaller than $A$ because of the twist angle inhomogeneity and the absorption is expected to be only a few percent, implying that the $\mu$ calculated here serves as an upper limit. Having fixed $\mu < 1$, the majority of pulses incident in the area $A$ carry either 0 or 1 photon and the probability of a pulse carrying 2 photons is negligible.

Supplementary Figure S12a illustrates the extracted detection efficiency (defined as the ratio of counts detected per second to photons incident per second in the area $A$) plotted against the laser repetition rate, revealing three distinct regimes. For $f_{RR} < 100$ Hz the detection efficiency decreases until it reaches a plateau which persists up to $f_{RR} \sim 30$ kHz. After this plateau, the detection efficiency abruptly drops. In the low repetition rate regime, the count rate is dominated by the dark counts: the rate at which the pulses carrying 1 photon are absorbed is lower than the dark count rate, resulting in a detection efficiency higher than the effective one. Within the range of repetition rates where we observe a plateau, the detection efficiency remains unaffected by the time distance between the pulses, indicating that the absorbed photon rate is smaller than the detector recovery time. This rules out steady state heating from the laser source for average powers below $< 300$ aW/$\mu m^2$. From the inset in Supplementary Figure S12a we observe that in this range of powers the PCR scales linearly with the average number of photons absorbed per second, confirming that the MATBG is operating as a single-photon detector. The drop of detection efficiency observed at high repetition rates is instead attributed to a saturation of the MATBG detector count rate and consistently occurs at the same average powers as in the CW experiment ($> 300$ aW/$\mu m^2$). Subsequently, we fix the repetition rate to 5 kHz, where the detection efficiency is independent of $f_{RR}$, and change the average laser power to control the number of photons carried on average by each pulse. Supplementary Figure S12b demonstrates that when the mean photon number per pulse is less than 1, the count rate evolves linearly with $\mu$ for two distinct bias voltages over several orders of magnitude. This observation further validates the single-photon sensitivity under pulsed excitation.

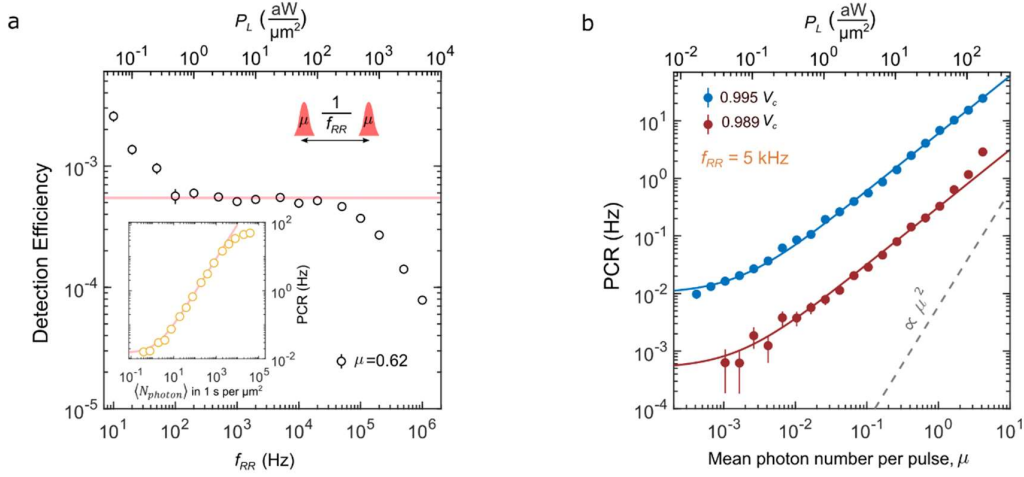

Supplementary Figure S12.| **Single-photon sensitivity with pulsed light excitation. (a)** Detection efficiency measured at fixed mean photon number per pulse $\mu = 0.62$ and different laser repetition rates, $f_{RR}$ for $V_{bias} = 0.995\, V_c$. Here the detection efficiency is defined as counts detected per second over photons incident per second in the area marked by the two voltage probes ($A \sim 16\,\mu m^2$). On the top x-axis the average incident power density $P_L$ corresponding to each $f_{RR}$. The solid line highlights the plateau in detection efficiency observed between 100 Hz to 30 kHz. Inset: photon count rate, PCR versus the average incident photon number $\langle N_{photon} \rangle$ in 1-s time window per $\mu m^2$. The solid line is a linear fit with an offset due to dark counts. **(b)** PCR versus $\mu$ for two different bias points at a fixed $f_{RR} = 5$ kHz. The solid lines are linear fits (with an offset due to dark counts), showing that the PCR evolves linearly with $\mu$.

## F. Additional photoresponse data of device A

For completeness we report additional photoresponse data measured on device A. In Supplementary Figure S13a we plot the PCR vs. $V_{bias}$ measured at various laser powers as in Fig. 3a of the main text but in linear scale. In the linear scale it is possible to see the sigmoidal shape with tendency to saturation. Specifically, in Supplementary Figure S13b, we plot the PCR vs. $P_L$ in correspondence of these saturation plateaus and show that they evolve linearly with laser power ruling out an artifact form the limited bandwidth. In Supplementary Figure S13c we plot the PCR vs. $P_L$ measured at a different bias point ($V_{bias} = 0.991\, V_c$) than the ones reported in Fig. 3 of the main text. Even at this $V_{bias}$ the MATBG detector shows single-photon sensitivity.

We also measure the PCR vs. $P_L$ measured at $T = 700$ mK and $V_{bias} = 0.996\, V_c$ (Supplementary Figure S14) and observe a linear scaling of the PCR with $P_L$, demonstrating single-photon sensitivity up to this temperature.

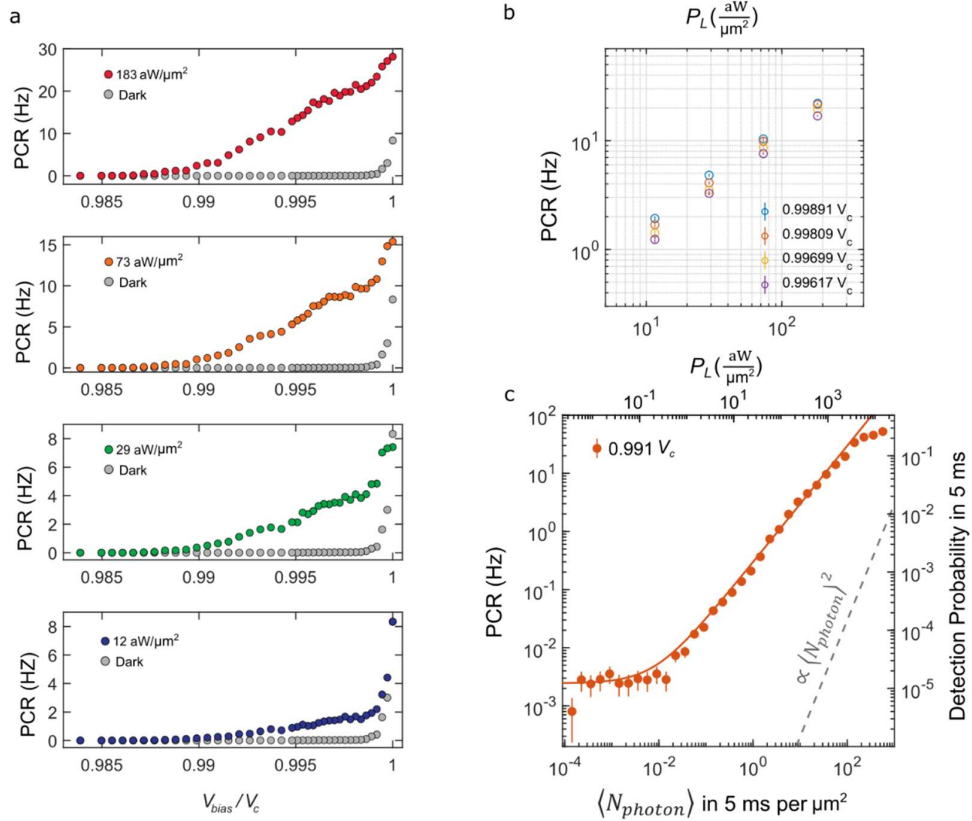

Supplementary Figure S13.| **Additional photoresponse data of device A. (a)** Extracted photon count rate (PCR) vs. $V_{bias}$ measured at various laser powers and plotted in linear scale. **(b)** PCR vs. $P_L$ in correspondence of the saturation plateaus ~ 0.997 $V_c$. **(c)** PCR vs. $P_L$ measured at $V_{bias}$ = 0.991 $V_c$. The MATBG detector shows single-photon sensitivity even at this bias point.

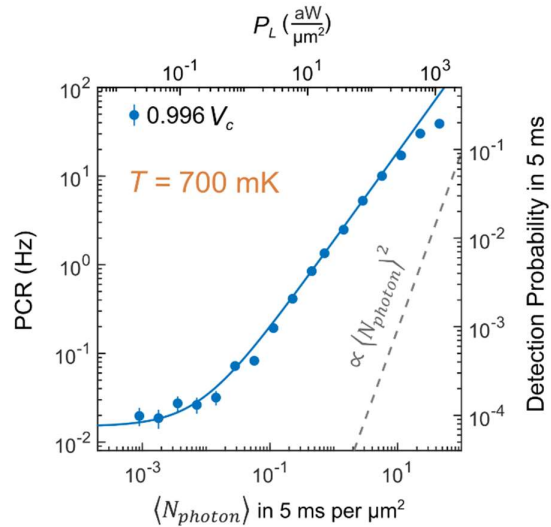

Supplementary Figure S14.| **Single-photon sensitivity at $T$ = 700 mK.** PCR vs. average incident photon number $\langle N_{photon} \rangle$ at $T$ = 700 mK and $V_{bias}$ = 0.996 $V_c$. The solid line is a linear fit (with an offset due to dark counts) demonstrating single-photon sensitivity up to 700 mK.

## G. Photovoltage generation and pulse shape

Here we show the pulse shapes measured for the three devices and discuss the origin of the photovoltage generated in MATBG devices. We argue that the photovoltage studied here is due to a complete breaking of the superconducting state upon absorption of a photon. In Supplementary Figure S15 we show the oscilloscope traces recorded while sweeping the bias voltage across the transition in both directions from the superconducting to normal state and vice-versa. By comparing them with the pulse shapes measured for device A, B and C (at $V_{bias}$ = 0.989 $V_c$, $V_{bias}$ = 0.991 $V_c$ and $V_{bias}$ = 0.9994 $V_c$ respectively) we notice that the voltage output induced by the photons matches the voltage generated by manually sweeping the device across the transition.

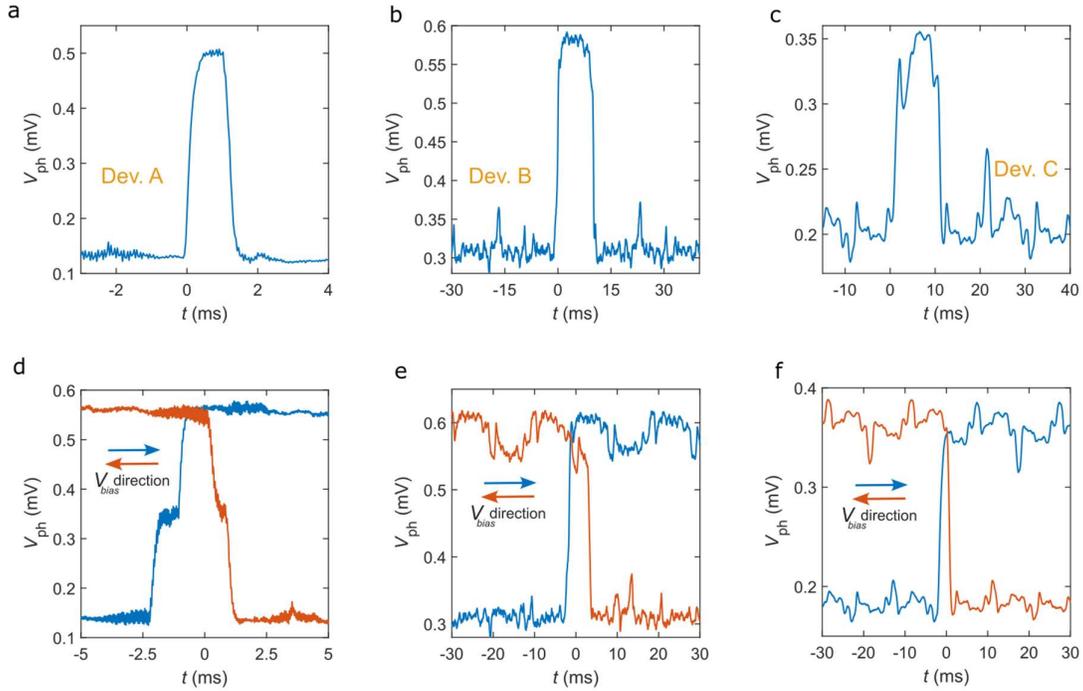

Supplementary Figure S15.| **Photovoltage generation and pulse shape for all devices. (a)-(c)** Typical pulse shape measured with a single-shot oscilloscope upon photo-absorption for device A, B and C at $V_{bias}$ = 0.989 $V_c$, $V_{bias}$ = 0.991 $V_c$ and $V_{bias}$ = 0.9994 $V_c$ respectively. **(d)-(f)** Single-shot oscilloscope traces recorded while sweeping the bias voltage across the transition in both directions from the superconducting to normal state (blue) and vice-versa (red).

## H. Photoresponse of device B

In this section we summarize the optoelectronic measurements performed on device B. As discussed in the transport characterization, device B features *I-V* curves very similar to device A but the superconducting state is not fully developed since it does not reach zero resistance (Supplementary Figure S4e). We attribute this to twist angle inhomogeneity[12]. In this device we measure the photovoltage time traces using the exact same setup and circuit presented above and also observe voltage spikes which increase as we increase the incident laser power. However, we notice that in this device the dark count rate is just slightly lower than the PCR with illumination. In addition, in this case the laser powers needed to measure an increase of PCR are considerably higher than the ones used for device A (see Supplementary Figure S17). We believe that both these observations can be ascribable to the presence of a thick graphite top gate (~ 7 nm) which is absent in device A. The additional graphite layer significantly absorbs radiation and acts as noise source for the MATBG detector. In light of these observations we are inclined to conclude that it is better to fabricate device which have a single graphite back-gate with thickness ~ 3 nm.

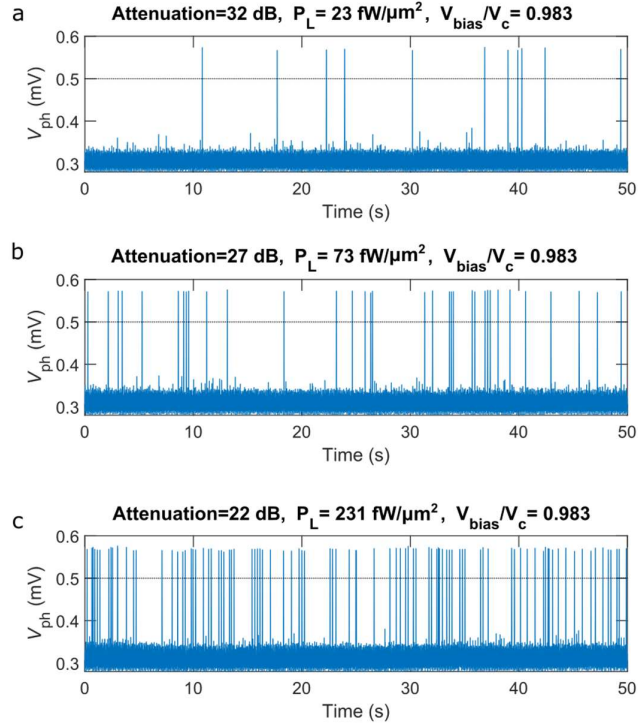

Supplementary Figure S16.| **Raw photovoltage time traces at different laser powers for device B. (a)-(c)** Raw photovoltage time traces measured over time for 3 laser powers. The black dashed line represents the threshold for counting the clicks. If the generated photovoltage is $V_{ph}$ > 0.5 mV, we register a count in the detector.

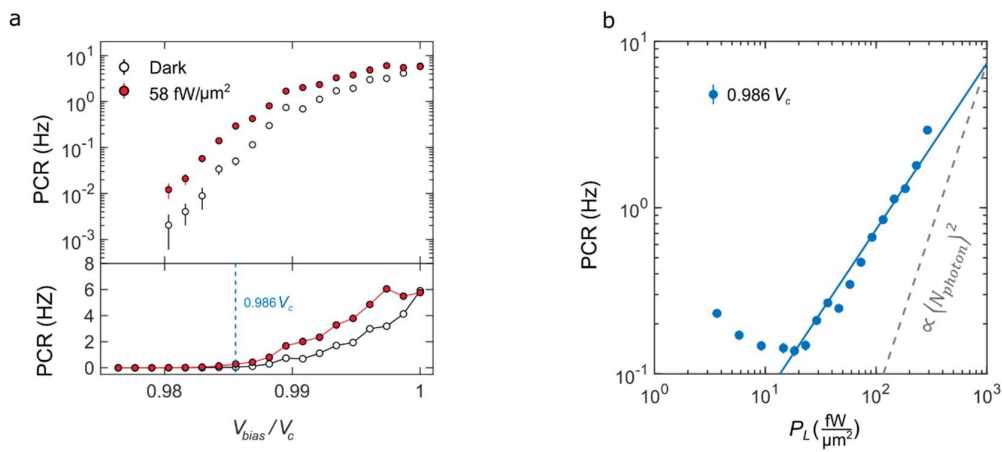

Supplementary Figure S17.| **Photoresponse of device B. (a)** Photon count rate (PCR) vs. $V_{bias}$ measured with and without 1550 nm laser-illumination. **(b)** Extracted photon count rate PCR versus laser power for the bias points indicated by the vertical dashed line in **(b)**. The solid line is a linear fit.

## I. Photoresponse of device C

In this section we summarize the optoelectronic measurements performed on device C. Device C features different *I-V* characteristics than device A and B because it has a smaller hysteresis loop (~ 2-3 nA). In addition, while device A and B feature a sharp transition from superconducting to normal state (in the switching the resistance changes ~ 9 kΩ and ~ 40 kΩ respectively), device C shows a non-linear behavior close to $I_c$ which results in a smooth transition and makes the MATBG detector unstable. In light of these considerations we argue that it is important to use MATBG devices in which the superconducting transition is as sharp as possible and that the difference in resistance between superconducting to normal state is large. In device C we were able to record some light-induced switching events as shown in Supplementary Figure S18. However, the clicks are not uncorrelated as expected for the photon shot noise coming from a highly attenuated laser source and the detector is unstable. In Supplementary Figure S18 we summarize the statistics performed on these photovoltage time traces. The count rate increases with laser power but the powers used in this experiment are considerably higher than the ones used for device A.

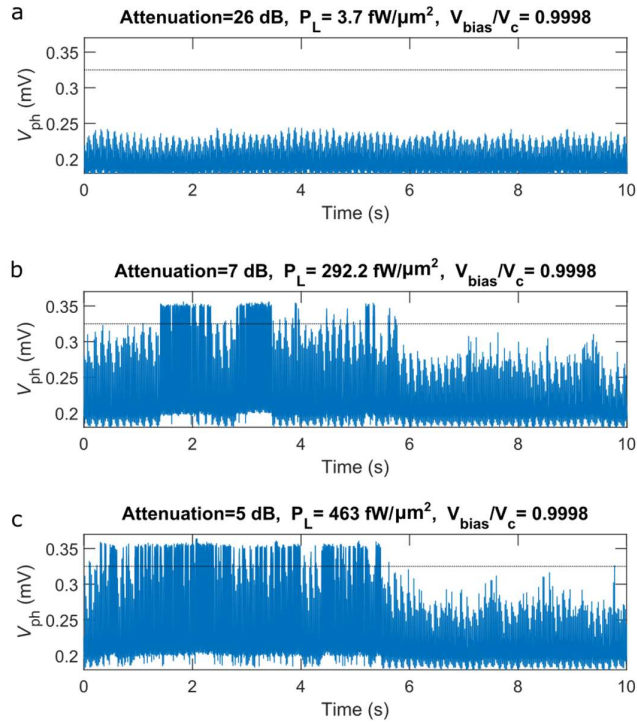

Supplementary Figure S18.| **Raw photovoltage time traces at different laser powers for device C.** **(a-c)** Raw photovoltage time traces measured over time for 3 laser powers. The black dashed line represents the threshold for counting the clicks. If the generated photovoltage is $V_{ph} > 0.325$ mV, we register a count in the detector.

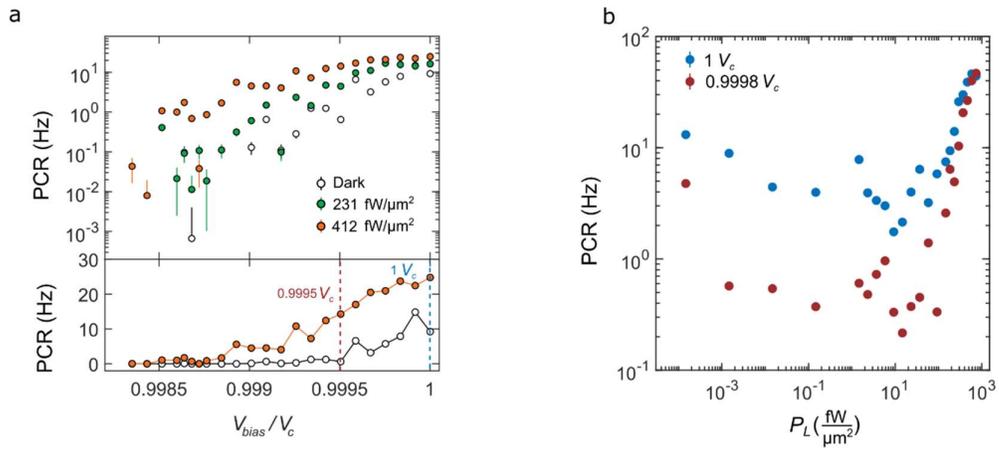

Supplementary Figure S19.| **Photoresponse of device C. (a)** Photon count rate (PCR) vs. $V_{bias}$ measured with and without 1550 nm laser-illumination. **(b)** Extracted photon count rate PCR versus laser power for the bias points indicated by the vertical dashed lines in **(b)**.